\documentstyle [12pt] {article}
\newcommand{\h}[1]{\mathop{\lambda}\limits_{#1}\ \!\!\!}

\newcommand{\edf}{\ {\mathop{=}\limits^{\rm def}}\ }

\newcommand{\al}{\alpha}

\begin{document}
\begin{center}
 \bf {QUANTUM PROPERTIES OF A GENERAL PATH DEVIATION EQUATION IN THE PAP-GEOMETRY}
\end{center}
\begin{center}
\bf{M.I.Wanas\footnote{Astronomy Department, Faculty of Science,
Cairo University, Giza, Egypt\\E-mail:wanas@frcu.eun.eg}}
\end{center}
\begin{center}
\abstract{ A path deviation equation in the Parameterized Absolute
Parallelism (PAP) geometry is derived. This equation includes
curvature and torsion terms. These terms are found to be naturally
quantized. The equation represents the deviation from a general
path equation, in the PAP-geometry, derived by the author in a
previous work. It is shown that, as the effect of the torsion, on
the deviation, increases, the effect of the curvature decreases.
It is also shown that the general path deviation equation can be
reduced to the geodesic deviation equation if PAP-geometry becomes
Riemannian. The equation can be used to study the deviation from
the trajectories of spinning elementary particles. }

\end{center}
\section{Introduction}
 In the context of geometerization of physics gravity, as a physical
 interaction, is well studied and understood . Unfortunately,
 other interactions are not understood in this context, so far, but
 they are well described in the framework of the quantization
 philosophy. There is no satisfactory descriptions of
 gravity in the context of quantization. In general, there is no
 unified description, for all known physical interaction. This
 represents the main problem facing the scheme of unification of
 physics. Several authors have tackled this problem, very early in
 the 20th century, in the hope of giving a solution, or at least
 to find an avenue for a solution, of this problem. In the first
 half of the 20th century, the attempts of the authors, to solve
 this problem, were in the framework of geometerization philosophy
 (e.g. [1], [2], [3]). After the apparent
 failure of these attempts, authors rushed, in the second half of
 the 20th century, to the opposite direction especially after the
 success of unifying electromagnetic and weak interactions
 (e.g. [4], [5], [6]) in the context of the
 quantization philosophy. The success in this direction is not
 continuous, since quantization of gravity is still beyond the
 reach of researchers. It seems that a new idea or philosophy
 is needed to solve this problem.

 Recently, the author and his collaborators [7], [8] have discovered that
 geometries with non-vanishing torsion possess some quantum
 properties. These properties appeared very naturally without
 applying any known quantization scheme. The first appearance of
 these properties is in the torsion term of a new set of path
 equations. In absolute parallelism (AP) geometry, this set can be
 represented by the following general equation
 $$\frac{d^{2}x^{\mu}}{d
 s^{2}}+a\{^{\mu}_{\al\beta}\}\frac{dx^{\al}}{d
 s}\frac{dx^{\beta}}{d
 s}=-b\Lambda^{.~.~\mu}_{(\al\beta)}\frac{dx^{\al}}{ds}\frac{dx^{\beta}}{ds},\eqno{(1.1)}$$
 where $\{~\}$ is the Christoffel symbol of the second type and
 $\Lambda^{\al}_{.~\mu\nu}$ is the torsion of the AP-space
 (cf. [9]). The coefficient $(a)$, of Christoffel symbol term,
 is always one in this set, while the coefficient $(b)$, of the
 torsion term, takes only the values $0,\frac{1}{2},1$ in the new
 set of the path equations (1.1). This is tempting to believe that
 paths in AP-geometry are naturally quantized. In the present work,
 an equation is said to be quantized if one or more of its terms
 do not take continuous values. More precisely, if the coefficients
 of such terms have jumping values, with constant step, the equation
 is said to be quantized or has some quantum properties. The set of
 equations, mentioned above, has been generalized and parameterized [10].
 The parameterized path equation is suggested to represent
 trajectories of spinning elementary particles in a back ground
 gravitational field. The torsion term, in this equation, is
 suggested to represent a type of interaction between the torsion
 of the back ground field and the quantum spin of the moving
 particle. Experimental and observational evidences for the
 existence of this interactions are examined [11], [12].
 Several applications and examinations of
 the parameterized path equations are carried out [13], [14].

 The study of the quantum properties, appeared in the AP-geometry,
 led to the development of a version of AP-geometry called the
 Parameterized Absolute Parallelism (PAP) geometry [15], [16], [17]
  in which torsion and curvature are
 simultaneously non-vanishing objects. The aim of the present work
 is to explore the quantum properties of path deviation
 equation in the PAP-geometry. For this reason, in the next
 section we give a brief review of the PAP-geometry. In section 3
 a new equation of path deviation, analogous to the
 equation of geodesic deviation of Riemannian geometry, is derived
 in PAP-geometry. The work is discussed and concluded in section
 4.

 \section{Parameterized Absolute Parallelism Geometry}
Since the Parameterized Absolute Parallelism (PAP) geometry is a
developed version of the (AP) geometry, I am going first to give a
brief idea about the AP-space. The AP-space is an n-dimensional
manifold, each point of which is characterized by a set of
n-linearly independent variables $x^{\nu}(\nu=1,2,.......,n)$, and
at each point we define a set of n-linearly independent
contravariant vectors $\h{i}^{\mu}~~(i=1,2,3,.....,n$ denotes the
vector number, and $\mu=1,2,3,......,n$ denotes the coordinate
components)\footnote{ In this paper, we are going to use Latin
indices to denote vector numbers and Greek indices to denote
coordinate components. Summation convention is used whatever the
location of indices. We are not going to raise or lower Latin
indices.}. The covariant form of these vectors, $\h{i}_{\mu} $,
are defined such that,
$$\h{i}^{\mu}\h{i}_{\nu}=\delta^{\mu}_{\nu},\eqno{(2.1)}$$
$$\h{i}^{\mu}\h{j}_{\mu}=\delta_{ij}.\eqno{(2.2)}$$ Using these
vectors, one can define the following second order symmetric
tensors
$$g_{\mu\nu}\edf\h{i}_{\mu}\h{i}_{\nu},\eqno{(2.3)}$$
$$g^{\mu\nu}\edf\h{i}^{\mu}\h{i}^{\nu}.\eqno{(2.4)}$$ The tensor
given by (2.3) can be used as a metric tensor of the Riemannian
space, associated with the AP-space, when needed. The condition of
Absolute Parallelism,
$$\h{i}_{\stackrel{\mu}{+}|\nu}=0,\eqno{(2.5)}$$ implies the
definition of the non-symmetric connection,
$$\Gamma^{\al}_{.~\mu\nu}\edf\h{i}^{\al}\h{i}_{\mu,\nu}.\eqno{(2.6)}$$
Since this connection is non-symmetric, one can define three types
of tensor  derivatives. The first is defined by using
$\Gamma^{\al}_{.~\mu\nu}$, and is denoted by a stroke and a (+)
sign (like (2.5)). The second is defined by using the dual
connection
$\tilde{\Gamma}^{\al}_{.~\mu\nu}(=\Gamma^{\al}_{.~\nu\mu})$ and is
denoted by a stroke and a (-) sign, and the third is defined by
using the symmetric part $\Gamma^{\al}_{.~(\mu\nu)}$ and is
denoted by a stroke. Of course one can add to these 3-derivatives
a 4th one, the covariant derivative, by using Christoffel  symbol
defined using (2.3), (2.4) and denoted, as usual, by a semicolon
(;). The use of these derivatives was a direct reason for
discovering the quantum properties in the new set of paths (1.1)
[7].

The quantum properties discovered in the AP-geometry is due to the
fact that this geometry admits a number of affine connections.
Introducing more affine connections in the AP-geometry we always
get the same quantum properties [8]. So, in order to extend these
properties throughout the geometry and to be sure that these
properties are generic features of the AP-geometry, the author got
the idea of parameterizing the AP-geometry which gives a new
version called the PAP-geometry [10], [15]. This version of the
AP-geometry is characterized by a general connection derived by,
linearly, combining the above mentioned affine connections using
four different parameters. The resulting object is not in general
an affine connection, unless we impose a metricity condition,
$$g_{\mu\nu||\sigma}=0,\eqno{(2.7)}$$ which reduces the
4-parameters to one. The double stroke characterizes tensor
derivatives using the resulting general connection
$\nabla^{\al}_{.~\mu\nu}$. In other words, (2.7) cannot be
satisfied unless the general connection is given by,
$$\nabla^{\al}_{.~\mu\nu}=\{^{\al}_{\mu\nu}\}+~b~\gamma^{\al}_{.~\mu\nu},\eqno{(2.8)}$$
where
$$\gamma^{\al}_{.~\mu\nu}\edf\h{i}^{\al}\h{i}_{\mu;\nu}.\eqno{(2.9)}$$
For some reasons [10], [15] the parameter $b$ is suggested to take
the value,
$$b~=~\frac{n}{2}\al\gamma,\eqno{(2.10)}$$ where $n$ is a natural
number, $\al(=\frac{e^{2}}{\hbar c}\simeq\frac{1}{137})$ is the
fine structure constant and $\gamma$ is a dimensionless parameter
to be fixed by experiment or observation. The type of geometry
constructed using the parameterized connection (2.8) is called the
\underline{Parameterized Absolute} \underline{Parallelism} (PAP)
geometry. In what follows we are going to give some of the main
features of this geometry.

It can be easily shown that (2.8) is an affine connection under
the group of general coordinate transformations. It is clear that
$\nabla^{\al}_{.~\mu\nu}$ is non-symmetric and consequently
implies a non-vanishing (parameterized) torsion defined by,
$${\Lambda^{*}}^{\al}_{.~\mu\nu}\edf\nabla^{\al}_{.~\mu\nu}-\nabla^{\al}_{.~\nu\mu}
= b \Lambda^{\al}_{.~\mu\nu},\eqno{(2.11)}$$ from which we can
define the parameterized vector,
$${C^{*}}_{\mu}\edf{\Lambda^{*}}^{\al}_{.~\mu\al}= b
C_{\mu}.\eqno{(2.12)}$$ The curvature tensor, corresponding to the
connection (2.8), can be defined as
$${B^{*}}^{\al}_{.~\mu\nu\sigma}\edf\nabla^{\al}_{.~\mu\sigma,\nu}-
\nabla^{\al}_{.~\mu\nu,\sigma}+\nabla^{\epsilon}_{.~\mu\sigma}
\nabla^{\al}_{.~\epsilon\nu}-\nabla^{\epsilon}_{.~\mu\nu}
\nabla^{\al}_{\epsilon\sigma}, \eqno{(2.13)}$$ which is clearly a
non-vanishing tensor. This curvature is a parameterized one which
can be written in the alternative form
$${B^{*}}^{\al}_{.~\mu\nu\sigma}=R^{\al}_{.~\mu\nu\sigma}+
b~{Q^{*}}^{\al}_{.~\mu\nu\sigma},\eqno{(2.14)}$$ where
$R^{\al}_{.~\mu\nu\sigma}$ is the Riemann-Christoffel curvature
tensor built from Christoffel symbol, as usual, and
$${Q^{*}}^{\al}_{.~\mu\nu\sigma}\edf\gamma^{\stackrel{\al}{+}}_{.~\stackrel{\mu}{+}
\stackrel{\sigma}{+}|\nu}-\gamma^{\stackrel{\al}{+}}_{.~\stackrel{\mu}{+}\stackrel
{\nu}{-}|\sigma}+
b~(\gamma^{i}_{.~\mu\nu}\gamma^{\al}_{.~i\sigma}-\gamma^{i}_{.~\mu\sigma}\gamma
^{\al}_{.~i\nu}),\eqno{(2.15)}$$ which is also a parameterized
tensor.

The general path equation, corresponding to the connection (2.8),
is also parameterized and is written in the form, $$\frac{d
Z^{\mu}}{d\tau}+\{^{\mu}_{\al\beta}\}~Z^{\al}~Z^{\beta}=-b~\Lambda^{.~.~\mu}_{(\al\beta)}
Z^{\al}~Z^{\beta},\eqno{(2.16)}$$ where $Z^{\mu}(=\frac{d
x^{\mu}}{d\tau})$ is a contravariant vector, tangent to the path
(2.16) and $\tau$ is the evolution parameter characterizing this
path.

It is to be noted that the general affine connection admitted by
PAP-geometry, (2.8), gives rise to, simultaneously, non-vanishing
torsion (2.11) and curvature (2.13). It can be shown [13] that
Riemannian geometry and the conventional AP-geometry can be
obtained from the PAP-geometry as special cases, corresponding to
$b=0$ and $b=1$, respectively.

\section{Equation of Path Deviation}
Bazanski [18], [19] has suggested a method to derive the geodesic
and geodesic deviation equation from one Lagrangian of the form,
$$L_{B}\edf g_{\mu\nu}U^{\mu}\frac{D\psi^{\nu}}{D
s},\eqno{(3.1)}$$ where $U^{\mu}$ is a unit tangent to the
geodesic,~ $\psi^{\nu}$ is the deviation vector and $s$ is the
evaluation parameter varying along the geodesic. The derivative of
the deviation vector, given in (3.1), can be related to the
covariant derivative using Christoffel symbol, as usually done, as
$$\frac{D\psi^{\nu}}{D
s}=\psi^{\nu}_{.~;\al}U^{\al}.\eqno{(3.2)}$$ Bazanski [18] has
shown that varying (3.1) with respect to the deviation vector, one
obtains the geodesic equation, while varying the same Lagrangian
(3.1) with respect to the tangent vector $U^{\mu}$, one can obtain
the geodesic deviation equation.

Now, in what follows we are going to examine the consequences of
replacing the covariant derivative in (3.2) by the tensor
derivative constructed using the parameterized connection (2.8),
in order to derive the corresponding path deviation equation. The
parameterized Lagrangian can be written in the form,
$$L\edf g_{\mu\nu}~Z^{\mu}\frac{D\chi^{\nu}}{D\tau},\eqno{(3.3)}$$
where,
$$\frac{D\chi^{\nu}}{D\tau}\edf\chi^{\nu}_{.~||\al}Z^{\al},\eqno{(3.4)}$$
and $Z^{\mu}(\edf\frac{dz^{\mu}}{d\tau})$ is the tangent to the
path (2.16), $\tau$ is its evolution parameter and $\chi^{\al}$ is
a vector representing the deviation from the general path (2.16).

Recalling the condition (2.7), we can write (3.3) in the form
$$L\edf
Z^{\mu}\frac{d\chi_{\mu}}{d\tau}-\nabla^{\al}_{.~\mu\nu}\chi_{\al}Z^{\mu}Z^{\nu}.\eqno{(3.5)}$$
It has been shown that performing the variation on (3.5) w.r.t.
the deviation vector, the author got [10] the general path
equation (2.16). Now performing the variation, on the same
Lagrangian, with respect to the vector $Z^{\sigma}$, we get the
following results:
$$\frac{\partial L}{\partial
Z^{\sigma}}=\frac{d\chi_{\sigma}}{d\tau}-\nabla^{\al}_{.~\sigma\nu}\chi_{\al}
Z^{\nu}-\nabla^{\al}_{.~\mu\sigma}\chi_{\al}Z^{\mu},\eqno{(3.6)}$$
\setcounter{equation}{6}
\begin{eqnarray}
\frac{d}{d\tau}(\frac{\partial L}{\partial
Z^{\sigma}})&=&\frac{d^{2}\chi_{\sigma}}{d\tau^{2}}-\nabla^{\al}_{.~\sigma\nu,\rho}
Z^{\rho}\chi_{\al}Z^{\nu}-\nabla^{\al}_{.~\sigma\nu}\frac{d\chi_{\al}}{d\tau}
Z^{\nu}\nonumber \\ &
&-\nabla^{\al}_{.~\sigma\nu}\chi_{\al}\frac{dZ^{\nu}}{d\tau}-\nabla^{\al}_{.~\mu\sigma,\rho}
Z^{\rho}\chi_{\al}Z^{\mu}\nonumber \\ & &
-\nabla^{\al}_{.~\mu\sigma}\frac{d\chi_{\al}}{d\tau}Z^{\mu}-\nabla^{\al}_{.~\mu\sigma}
\chi_{\al}\frac{dZ^{\mu}}{d\tau},\label{(3.7)}
\end{eqnarray}
 and $$\frac{\partial L}{\partial
 x^{\sigma}}=-\nabla^{\al}_{.~\mu\nu,\sigma}\chi_{\al}Z^{\mu}Z^{\nu}.\eqno{(3.8)}$$
 Now substituting from (3.7) and (3.8) into the Euler-Lagrange
 equation, $$\frac{d}{d\tau}(\frac{\partial L}{\partial
 Z^{\sigma}})-\frac{\partial L}{\partial
 x^{\sigma}}=0,\eqno{(3.9)}$$ and making use of the general path
 equation (2.16), we get after somewhat long manipulations,
 $$\frac{D^{2}\chi_{\sigma}}{D\tau^{2}}+{B^{*}}^{\al}_{.~\mu\sigma\nu}
 Z^{\mu}Z^{\nu}\chi_{\al}=-b~\Lambda^{\al}_{.~\sigma\nu}\frac{D\chi_{\al}}
 {D\tau}Z^{\nu},\eqno{(3.10)}$$ where
 ${B^{*}}^{\al}_{.~\mu\sigma\nu}$ is the parameterized curvature
 defined by (2.13). This is the parameterized path deviation equation, in
 which the torsion term on its R.H.S. is explicitly parameterized
 while the curvature term is implicitly parameterized via (2.14).

 \section{Discussion and Concluding Remarks}
 In what follows, we are going to show that the general path
 deviation equation will reduce to the geodesic deviation equation
 in the case of Riemannian geometry. Let the
 parameterized path deviation equation (3.10) be written,
 using (2.14), in the
 form:$$\frac{D^{2}\chi_{\sigma}}{D\tau^{2}}+(R^{\al}_{.~\mu\sigma\nu}+b~{Q^{*}}
 ^{\al}_{.~\mu\sigma\nu})Z^{\mu}\chi_{\al}Z^{\nu}=-b~\Lambda^{\al}_{.~\sigma\nu}
 \frac{D\chi_{\al}}{D\tau}Z^{\nu}.\eqno{(4.1)}$$ The case of
 Riemannian geometry could be achieved upon taking $b=0$ [10] then we get
 $(\tau\longrightarrow s, \chi_{\sigma}\longrightarrow
 \psi_{\sigma}, Z^{\mu}\longrightarrow U^{\mu})$
 $$\frac{D^{2}\psi_{\sigma}}{D
 s^{2}}+R^{\al}_{.~\mu\sigma\nu}~U^{\mu}\psi_{\al}~U^{\nu}=0.\eqno{(4.2)}$$
 Now using the symmetries of Riemann-Christoffel curvature
 tensor, $R^{\al}_{.~\mu\sigma\nu}$, and recalling that raising
 and lowering tensor indices commute with covariant
 differential operator we can write,
 $$\frac{D^{2}\psi^{\sigma}}{D
 s^{2}}=R^{\sigma}_{.~\nu\mu\al}U^{\mu}\psi^{\al}U^{\nu},\eqno{(4.3)}$$
 which is the well known geodesic deviation equation of Riemannian
 geometry.

 On the other hand, the case of the conventional
 AP-geometry corresponds to the value of the parameter $b=1$
  [10]. Substituting this value into
 equation (4.1) we get $(\tau\longrightarrow s^{+}, \chi_{\sigma}\longrightarrow
 \xi_{\sigma}, Z^{\mu}\longrightarrow V^{\mu})$
 $$\frac{D^{2}\xi_{\sigma}}{D
 {s^{+}}^{2}}+M^{\al}_{.~\mu\sigma\nu}V^{\mu}\xi_{\al}V^{\nu}=-\Lambda^{\al}
 _{.~\sigma\nu}\frac{D\xi_{\al}}{D s^{+}}V^{\nu},\eqno{(4.4)}$$
 where $M^{\al}_{.~\mu\sigma\nu}$ is the curvature tensor
 corresponding to the connection (2.6) and is given by,
 $$M^{\al}_{.~\mu\sigma\nu}\edf
 R^{\al}_{.~\mu\sigma\nu}+Q^{\al}_{.~\mu\sigma\nu},\eqno{(4.5)}$$
 where
 $$Q^{\al}_{.~\mu\sigma\nu}\edf\gamma^{\stackrel{\al}{+}}_{.~\stackrel{\mu}{+}
 \stackrel{\nu}{+}|\sigma}-\gamma^{\stackrel{\al}{+}}_{.~\stackrel{\mu}{+}
 \stackrel{\sigma}{-}|\nu}+\gamma^{\epsilon}_{.~\mu\sigma}\gamma^{\al}_{.~\epsilon\nu}
 -\gamma^{\epsilon}_{.~\mu\nu}\gamma^{\al}_{.~\sigma\epsilon}.\eqno{(4.6)}$$
 It is well known (cf. [20]) that the tensor (4.5) vanishes
 identically because of the AP-condition (2.5). Thus the equation
 of path deviation in the conventional AP-geometry (4.4), can be
 written in the form:
 $$\frac{D^{2}\xi_{\sigma}}{D{s^{+}}^{2}}=-\Lambda^{\al}_{.~\sigma\nu}\frac{D\xi_{\al}}
 {D s^{+}}V^{\nu}.\eqno{(4.7)}$$ Recalling that, because of the
 AP- condition (2.5) and definitions (2.3) and (2.4), raising and
 lowering tensor indices commute with the tensor differential
 operator using the connection (2.6). Then (4.7) can be written as
 $$\frac{D^{2}\xi^{\sigma}}{D{s^{+}}^{2}}=\Lambda^{.~.~\sigma}_{\al\nu}
\frac{D\xi^{\al}}{D s^{+}}V^{\nu},\eqno{(4.8)}$$ which is the
equation of the path deviation of the conventional AP-geometry
[21]. It is to be considered that equations (4.3) and (4.8)
represents the two extremes of the parameterized equation (3.10).
In the case of (4.3) the deviation depends only on the curvature
tensor, since the torsion is vanishing in Riemannian geometry;
while in the case of (4.8) the deviation depends only on the
torsion, since the curvature corresponding to (2.6) vanishes.
Between these two extremes, the deviation depends on both
curvature and torsion with the quantum steps of the parameter $b$
given by (2.10). The increase of the dependence on the torsion
corresponds to a decrease of the dependence on the curvature and
vice-versa. This shows how wide is the PAP-geometry [13], [17].

The quantum properties of the AP-geometry appeared, very naturally
i.e. without imposing any condition, in the torsion term of the
path equations [7]. The parameterization of this geometry is
carried out to extend these properties to other geometric objects
[10], [15]. Thus parameterization means, in some sense,
quantization. From the application point of view, the value of the
parameter $(b)$, given by (2.10), depends on the spin of the
moving particle [10]. Spin-less particle $(n=0\longrightarrow
b=0)$ will move along a geodesic and the deviation will depend on
the curvature only, as clear from (4.1). So, such particle will
feel the full curvature of the space, $R^{\al}_{.\mu\nu\sigma}$,
without any interaction with torsion. Consequently, one can
conclude that the dynamics of such a particle can be studied in
the context of Riemannian geometry. This is not because Riemannian
geometry has a vanishing torsion, but because of the vanishing
spin of such a particle.

One may jump to the conclusion that the quantum properties
appeared in the path deviation (3.10) are direct consequences of
parameterization scheme carried out in the AP-geometry. In general
this is not true [21]. It is shown that path deviation equations,
in AP-geometry, are naturally quantized, from a qualitative point
of view. The PAP-geometry is a modified version of the AP-geometry
constructed in order to build a geometry that can be used to
study, quantitatively and qualitatively, physical phenomena with
quantum properties.

\section*{References}

{[1] Einstein, A. (1930) Math. Annal. {\bf 102}, 685.} \\ {[2]
Schrodinger, E. (1947) Proc. Roy. Irish Acad., A {\bf 51}, 163} \\
{[3] Weyl, H. (1950) Phys. Rev., {\bf 2}, 699.} \\ {[4] Weinberg,
S. (1967) Phys. Rev., {\bf 2}, 699.} \\ {[5] Salam, A. (1968) In
{\it "Elementary Particle Theory"} ed. Svartholm, W., Almquist

and Wiskell. Stockholm.} \\ {[6] Glashow, S. L. (1961) Nucl. Phys.
{\bf 22}, 579.} \\ {[7] Wanas, M.I., Melek, M. and Kahil, M.E.
(1995) Astrophys. Space Sci. {\bf 228}, 273},

gr-qc/0207113. \\ {[8] Wanas, M.I.
and Kahil, M.E. (1999) Gen. Rel. Grav., {\bf 31}, 1921; gr-qc/9912007.}  \\
{[9] Wanas, M.I. (2001) Proc. 11th Nat.Conf. {\it "Fainsler,
Lagrange and Hamilton

Geometry"} Bacau 2000,Stud.Cercet.Stiin.Ser. 10, 297, eds
V.Blanute and Gh.Neagu ;

   gr-qc/0209050.} \\
{[10] Wanas, M.I.
(1998) Astrophys. Space Sci. {\bf 258}, 237; gr-qc/9904019.} \\
{[11] Wanas, M.I.,
Melek, M. and Kahil, M.E. (2000) Gravit. Cosmol. {\bf 6}, 319;

 gr-qc/9812085.} \\
{[12] Wanas, M.I., Melek, M. and Kahil, M.E. (2002) Proc. MG IX,
vol.{\bf 2}, 1100;

gr-qc/0306086.} \\ {[13] Wanas ,M.I. (2003) Algebras, Groups and
Geometries, {\bf 20}, 345a.} \\ {[14] Sousa,
A.A. and Maluf, J.W. (2004) Gen.Rel. Grav. {\bf 36}, 967; gr-qc/0310131.} \\
{[15] Wanas M.I. (2000) Turk. J. Phys. {\bf 24}, 473; gr-qc/0010099.} \\
{[16] Wanas, M.I. (2002) Proc. MG IX, vol. {\bf 2}, 1303.} \\
{[17] Wanas, M.I. (2003) Proc. XXV Int. Workshop {\it "Fundamental
Problems of H.E.Physics"},

 315}, held in Protvino, Russia, 22-28 June 2002. \\ {[18] Bazanski, S.L. (1977) Ann. Instit.
H. Poincare, {\bf A27}, 145.} \\ {[19] Bazanski, S.L. (1989) J.
Math. Phys. {\bf 30}, 1018.} \\ {[20] M\o ller, C. (1961) Math.
Fys. Skr. Dan. Vid. Selsk, {\bf 1}, 1.} \\ {[21] Wanas, M.I. and
Kahil, M.E. (2004) to appear in Proc. MG X}.

\end{document}